\documentstyle[12pt,epsfig]{article}
\newlength{\dinwidth}
\newlength{\dinmargin}
\let \captionsize=\small

\begin{document}
\title{Spin dependent structure function $ g_{2}(x)$ in quark-parton model.
Possible interpretation and numerical estimates }
 \author{N.L.Ter-Isaakyan}
\date{Yerevan Physics Institute}
\maketitle{}
\abstract{It is shown that in the special infinite momentum frame where photon
has pure transverse components at $P\rightarrow\infty$ the spin-dependent
deep inelastic structure function $ g_{2}(x)$ has a reasonable
interpretation in terms of quark-parton wave functions, whereas in the
conventional frame where photon has pure z-component the parton model
fails for $ g_{2}(x)$.The spin dependent structure functions $ g_{1}(x)$
and $ g_{2}(x)$ have been calculated in the relativistic quark model
constructed in such frame. The results indicate significant twist-3
contribution.}

 \section{Introduction}
Experiments designed to measure the spin-dependent structure functions of
the nucleon are now being performed (see, e.g.,\cite{1}). As it is well
known the structure function $g_{2}(x)$ has a transparent interpretation
in the quark parton model (see, e.g.\cite{2,3} ), whereas the parton
interpretation of $ g_{2}(x)$ faces serious difficulties , which are
connected with the fact that this function turned out to be zero for a
free quark (see e.g.,\cite{3} ) and, hence, cannot be presented by its value
on a free quark averaged over the proper probabilities of parton
distributions. Therefore, the nonzero value of $ g_{2}(x)$ can be obtained
if we take into account parton interactions (or parton off-shellness). But
the results of such calculations depend on the coordinate system (see
below) and turn out to be physically unreasonable.
So, in the conventional frame where the proton has pure z-component
\cite{2,3}
\begin{equation}
P_{\mu}=(E,P,0,0),~~ q_{\mu}=(0,-2Px,0,0),
\label{BF}
\end{equation}
the value of $g_{1}(x)$+$g_{2}(x)$ vanishes for massless quarks and the
Burkhardt - Cottingham sum rule \cite{4},
\begin{equation}
\int_{0}^{1} {g_{2}(x) dx} = 0
\label{2}
\end{equation}
is not fulfilled.

It is convinient to analize the parton model considering the old fashioned
perturbation theory diagrams in the infinite momentun frame (IMF).
  In general, when we want to
take into account parton interactions, in addition with naive parton model
diagram of Fig. \ref{koko}, the diagrams of fig.2 which contain
$q\overline{q}$-pairs creation or annihilation must be also taken into
account.

\begin{figure}[htb]  \centering \unitlength 1mm
\begin{picture}(50,50)
\put(-34,42){\begin{picture}(50,50)
\includegraphics{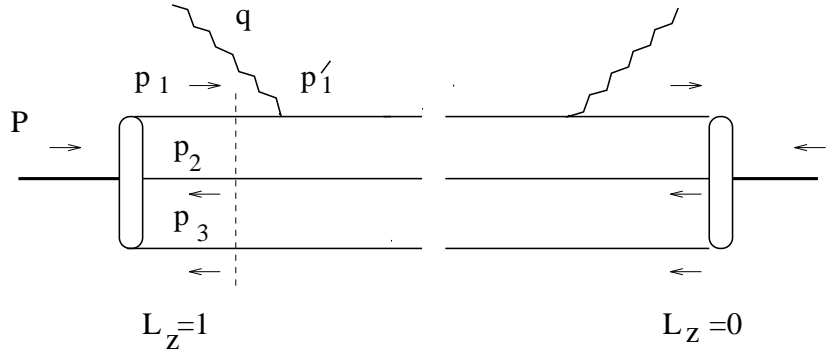}
\end{picture}}
\end{picture}
\caption{The naive parton model diagram of deep
inelastic scattering. The  arrows  denote  nucleon  and  guarks
helicities for helicity-flip  amplitude   which   determine structure
function $g_{2}(x)$ }
\label{kok}
\end{figure}
\begin{figure}[htb]  \centering \unitlength 1mm
\begin{picture}(50,95)
\put(-34,92){\begin{picture}(50,95)
\includegraphics{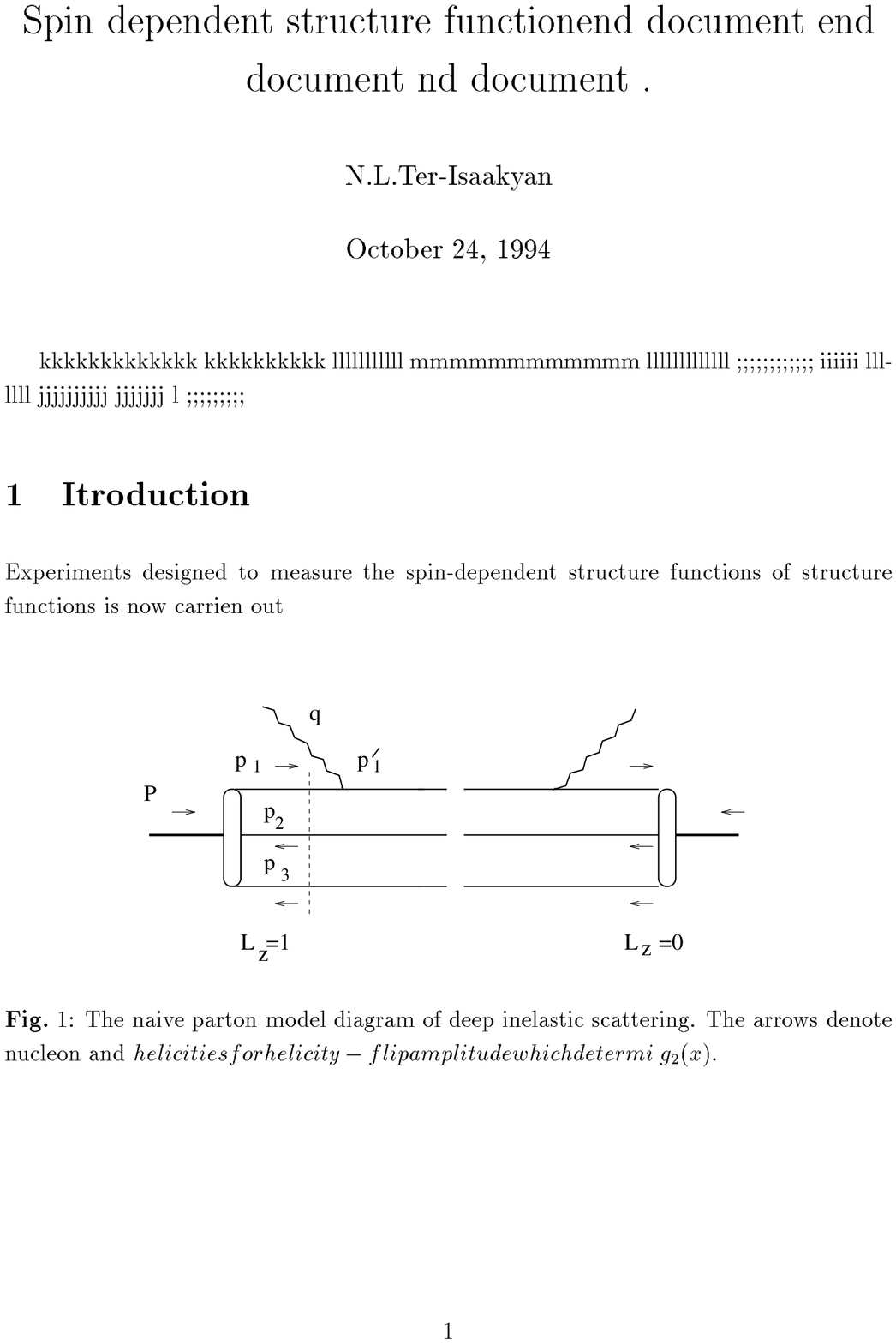}
\end{picture}}
\end{picture}
\caption{\captionsize The diagrams, which could violate the parton model
for $g_{2}(x)$.}
\label{o}
\end{figure}
\begin{figure}[htb]  \centering \unitlength 1mm
\begin{picture}(50,50)
\put(-50,43){\begin{picture}(50,50)
\includegraphics{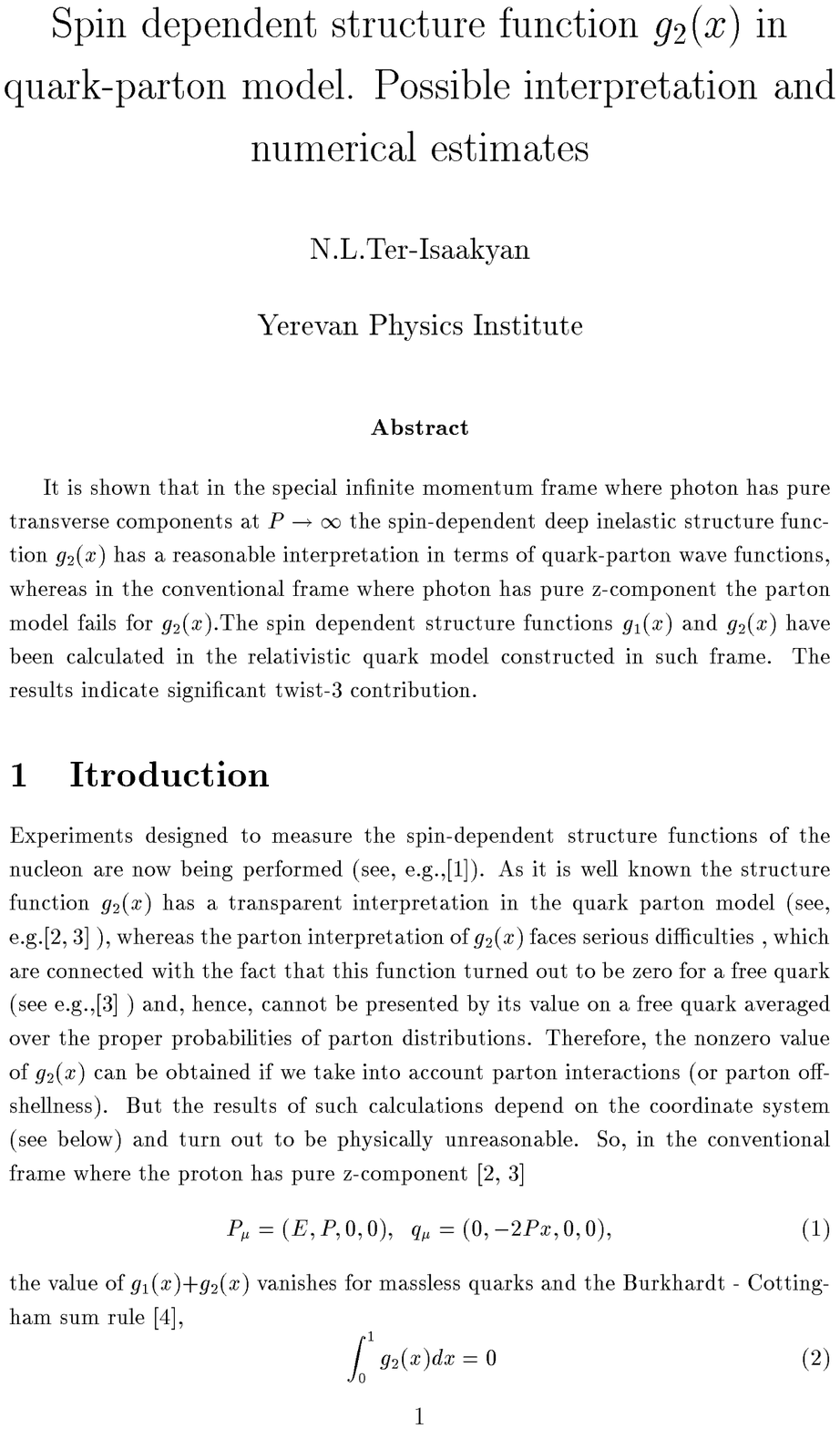}
\end{picture}}
\end{picture}
\caption{\captionsize The diagrams, which determine the contribution to
      $g_{2}(x)$ of
      Compton amplitude with different number of  partons  in  the
      initial and final nucleon wave functions.}
\label{koko}
\end{figure}

 It is shown in this paper that such diagrams may contribute to
$g_{2}(x$) and violate the validity of parton model. We show that only in
special IMF it is possible to get reed of such diagrams and to represent $
g_{2}(x)$ in terms of quark-parton infinite momentum wave functions.  I
find that there are two phenomenologically independent contributions to $
g_{2}(x)$.  The first contribution is determined by the diagrams with the
different values of quark orbital angular momentum projections in the
initial and final nucleon wave functions (fig.1)\footnote{For simplicity,
all diagrams are given for three quark state}. The second contributionis
determined by Compton amplitudes with different number of gluons in the
initial and final nucleon wave functions (fig.3). In the field theory these
two contributions may be apparently connected.

The similar interpretation of $g_{2}(x)$ were discussed in Ref.5 in the
framework of operator product expansion on "light cone".

In the second part of this paper I calculate the structure functions
$g_{1}(x)$ and $g_{2}(x)$ in the relativistic quark model of ref.6. The
results agree with the bag model calculations \cite{7} and with QCD sum
rule result \cite {8,9} and indicate a significant twist-3 contribution
to $g_{2}(x)$ in the range of $x\leq 0.5$.

\section{Quark-parton interpretation of $g_{2}(x)$}
  We start from the standart definition of the hadronic tensor:
$$\frac{M}{4\pi}W_{\mu\nu}=\sum_{X}(2\pi)^{4}\langle p,s|J_{\mu}|X\rangle
\langle X|J_{\nu}|p,s\rangle\delta(P+q-P_{X})=$$

\begin{equation}
P_{\mu}P_{\nu}W_{2}-M^{2}W_{1}g_{\mu\nu}+
iM\epsilon_{\mu\nu\lambda\sigma}q^{\lambda}[M^{2}s^{\sigma}G_{1}+
(Pqs^{\sigma}-sqP^{\sigma})G_{2}]
\label{3}
\end{equation}
(disregarding terms proportional to $q_{\mu}$ or $q_{\nu}$ ), where
$W_{1}$, $W_{2}$, $G_{1}$ and $G_{2}$ are functions of $q^{2}$ and $\nu$=
$Pq/M$. The nucleons are supposed in the same spin state described by
$s^{\mu}$. In the coordinate system (\ref{BF}) the spin average structure
functions $F_{1}(x)$=$MW_{1}(x)$ and
$F_{2}(x)$=$\nu W_{2}(x)$=$2xF_{2}(x)$ are
determined by symmetric part of the hadronic tensor $W_{ij}^{s}$
$(i,j=1,2)$.
The spin dependent structure functions may be expressed through
antisymmetric parts of $W_{ij}$ and $W_{i0}$ $(i=1,2)$ as follows:

$$\frac{1}{2\pi}W_{ij}^{a}=2i \epsilon_{ij} g_{1}(x) \frac{s_{0}}{2P}$$

\begin{equation}
\frac{1}{2\pi}W_{i0}^{a}=i\epsilon_{ij}s_{j}(g_{1}(x)+g_{2}(x))
\frac{2M}{P}
\label{4}
\end{equation}
where the functions $g_{1}$ = $M^{2}\nu G_{1}$ and  $g_{2}$ = $M\nu^{2}
G_{2}$ scales in the Bjorken limit.For quark momenta (which are defined
on diagrams) we introduce the  standart parameterizations:

$$\vec{p}_{1}=x_{1}\vec{P}+\vec{p}_{1\perp},~~
\vec{p}'_{1}=-x_{1}\vec{P}+\vec{p}_{1\perp},~~
\vec{p}''_{1}=-x_{1}\vec{P}-\vec{p}_{1\perp},$$

\begin{equation}
{\vec{p}}_{1\perp}\vec{P}=0
\label{5}
\end{equation}

In the coordinate system (1) the vertices of photon  interactions
with quartks (on fig.1) and with $q\bar{q}$-pairs (on fig.2) at
$P\rightarrow \infty$ behave as follows (i,j=1,2):

$$\bar{u}(p'_{1})\gamma_{i} u(p_{1})=2Px_{1}\sigma_{i}
\sigma_{3},~~~
\bar{u}(p'_{1})\gamma_{0} u(p_{1})=2(m+i\epsilon_{ik}
\sigma_{i} p_{1k}),$$

\begin{equation}
\bar{u}(p'_{1})\gamma_{i} v(p''_{1})=2(m\sigma_{i}+
i\epsilon_{ij}p_{1j})\sigma_{2},~~
\bar{u}(p'_{1})\gamma_{0} v(p''_{1})=2Px_{1}\sigma_{3}
\sigma_{2}.
\label{6}
\end{equation}
The amplitude of  antiquark  interaction  with  nucleon  also  may
behave as  P.  Hence,  the  large  energy  denominators
corresponding to  dashed  lines  on  diagrams  of  fig.2  may  be
compensated for $W_{i0}$   (but not  for  $W_{ij}$  )  and  these  diagrams
may contribute to ($g_{1}(x)$+$g_{2}(x)$). Thus, the naive  parton  model
fails for $g_{2}(x)$ in the coordinate system (1).
   Let us consider now the special  IMF  where  photon  have  pure
transverse component at $P\rightarrow \infty$:

\begin{equation}
P_{\mu}=(E,P,0,0),~~ q_{\mu}=(\frac{q_{\perp}^{2}}{4Px},
\frac{-q_{\perp}^{2}}{4Px},0,0).
\label{7}
\end{equation}
In this frame the structure functions  $g_{1}(x)$ and $g_{2}(x)$ are
expressed  through   anitisymmetric component  of   the   hadronic
tensor $W_{i0}^{a}$ in  the following form:

$$\frac{1}{4\pi}W_{i0}^{a}\frac{q_{\perp}^{2}}{x}=
2i\epsilon_{ij} q_{j}s_{0} g_{1}(x)$$

\begin{equation}
\frac{1}{4\pi} W_{i0}^{a} \frac{q_{\perp}^{4}}{x^{2}}=
2i\epsilon_{ij} q_{j} \vec{s} {\vec{g}}_{\perp}
Mg_{2}(x)
\label{8}
\end{equation}
We shall take longitudinally polarized nucleon to extract $g_{1}(x)$ and
transversely polarized nucleon in $\vec{q}_{\perp}$  direction to extract
 $g_{2}(x)$.  But it is more transparent physically to represent
$g_{2}(x)$ in
terms of helicity amplitudes.  In that language $g_{2}(x)$ corresponds to
nucleon helicity-flip Compton amplitude. In the IMF (7) the vertices of
photon interactions with quartks (on fig.1) and with $q\bar{q}$-pairs (on
fig.2) at $P\rightarrow \infty$ behave as follows (i=1,2):

$$\bar{u}(p'_{1})\gamma_{0} u(p_{1})=2Px_{1},~~
\bar{u}(p'_{1})\gamma_{i} u(p_{1})=2(p_{1i}+q_{i}+i\epsilon_{ik}q_{k}
\sigma_{3}),$$
\begin{equation}
\bar{u}(p'_{1})\gamma_{0} v(p''_{1})=\vec{\sigma}\vec{q}\sigma_{3}
\sigma_{2},~~~~~
\bar{u}(p'_{1})\gamma_{i} v(p''_{1})=-2Px_{1}\sigma_{i} \sigma_{2}.
\label{9}
\end{equation}

In  the scaling limit the energy $\delta$-function gives:

\begin{equation}
\delta(E+q_{0} -E_{X})=\frac{2Pxx_{1}}{q_{\perp}^{2}}
\delta[x_{1}-x(1+\frac{2\vec{q}\vec{p}_{1\perp}}{q_{\perp}^{2}})].
\label{10}
\end{equation}

The  contributions of diagrams of fig.2 to $W_{i0}$ do not vanish at
$P\rightarrow \infty$, this amplitude behaves as
$W_{i0} \sim P/q_{\perp}^{2}$  in the scaling  limit and these
diagrams could
vcontribute, in principle, to $g_{2}(x)$. But it is easy  to see that
$W_{i0}$ do not depend on $\vec{q}_{\perp}$   direction  and  hence
cannot contain  necessary structure: $\epsilon_{ij}q_{j}\vec{s}\vec{q}$.
That means that such diagrams do not actually contribute to $g_{2}(x)$
and its correct  value can be derived taking into account only diagram
of fig.1,  if  we compare, for instance, terms proportional to
$q_{i}q_{j}$  at both sides  of (8).

At first sight the diagram of fig.1 also do  not  contribute  to
$g_{2}(x)$  because  the  corresponding  value  of $W_{i0}$    has
a wrong $\vec{q}_{\perp}$ dependence and do not contain  spin-flip terms.
To
obtain  the nonzero value of $g_{2}(x)$ we have to take into account the
second term of the expansion of $\delta$ - functions argument at
$q_{\perp}^{2}\rightarrow \infty$ in (10).
   Finally, the contribution of  the  diagram  of  Fig.1   may  be
presented in the form:

$$g_{2}^{(1)}(x)=\frac{d}{dx}\bar{g}(x),$$

\begin{equation}
2M\bar{g}(x)=\sum_{r}\int d\Gamma^{n}\delta(x-x_{r})\sum_{s_{1},...s_{n}}
\Psi_{s_{i}}^{\ast\uparrow}(x_{i},\vec{p}_{i\perp})2s_{r}Q_{r}^{2}
\Psi_{s_{i}}^{\downarrow}(x_{i},\vec{p}_{i\perp})(p_{r\perp}^{x}+ip_{r\perp}^{y}),
\label{11}
\end{equation}
where $d\Gamma^{n}$ is n-particle phase space:

\begin{equation}
d \Gamma^{n}=\frac{1}{x_{n}} \sum_{i=1}^{n-1} \frac{dx_{i}
d\vec{p}_{i \perp}}{2x_{i}(2\pi)^{3}},
\label{12}
\end{equation}
$Q_{r}$  and $s_{r}$  denote charge and spin projection of active quark
along z direction. The energy denominators  are  included  into  nucleon
wave function:

\begin{equation}
\Psi_{s_{i}}(x_{i},\vec{p}_{i\perp})=\frac{\Gamma_{s_{1}...s_{n}}
(x_{1},\vec{p}_{1\perp}...x_{n},\vec{p}_{n\perp})}
{2P(E- \sum_{i} E_{i})}
\label{13}
\end{equation}
where $\Gamma_{s_{1}...s_{n}}(x_{1},\vec{p}_{1\perp}...x_{n},\vec{p}_{n\perp})$
is the nucleon-partons vertex.

The normalization condition of nucleon  wave  function (13) can be
fixed,  for  instance,  from   normalization  of  nucleon  electric
formfactor $F_{1}(Q^{2})$ at $Q^{2}$ =0 in  the quark parton  model
and  has  a following  form:

\begin{equation}
\int d \Gamma^{n} \sum_{s_{1},...s_{n}}
\Psi_{s_{i}}^{\ast}(x_{i},\vec{p}_{i\perp})
\Psi_{s_{i}}(x_{i},\vec{p}_{i\perp})=1
\label{14}
\end{equation}
(For more  details of deriving Egs.12-14 for three quark states see
ref.5). In the integral over transverse momentum in (11) only terms
which contains linear powers of transverse  momenta  in
the  final  or  initial  wave  functions  contribute.  Such   terms
can  arize  when  the  difference  of  angular   orbital   momentum
projections of initial and final states is equal to unity,
$\Delta \langle L_{z} \rangle$ =1; for such  states
the  sum of parton  helicities is not  equal  to  nucleon  helicity  and
hence nucleon spin-flip could take place as  it  is  shown  on  the
diagram (fig.1).
  The sum rule (2) will be fulfilled for (11) if $\bar{g}(1)$=
$\bar{g}(0)$=0.  The bound state wave  functions  vanish  at  x =0
 and  x =1   and  the QCD-evolution should not violate  the  sum  rule
 (2).  So,  if  we suppose that quark- antiquark sea  in  the  nucleon
arise  due  to QCD-evolution  and  at  low  resolution  scale  nucleon
can    be considered as a bound state of finite number of  constituents,
the sum rule (2) will be fulfilled for (11).

The structure function $g_{1}(x)$ in the same notations has a following
form:

\begin{equation}
2g_{1}(x)=\sum_{r} \int d \Gamma^{n} \delta(x-x_{r}) \sum_{s_{1},...s_{n}}
\Psi_{s_{i}}^{\ast \uparrow}(x_{i},\vec{p}_{i\perp}) 2s_{r}Q_{r}^{2}
\Psi_{s_{i}}^{\uparrow}(x_{i},\vec{p}_{i\perp}).
\label{15}
\end{equation}

Now consider the diagrams of fig.3. The momenta are  defined  on
the diagrams:

\begin{equation}
\vec{p}_{1}=x_{1}\vec{P}+\vec{p}_{1\perp},~~
\vec{k}_{1}=y_{1}\vec{P}+\vec{k}_{1\perp},~~
\vec{k}_{g}=y_{g}\vec{P}+\vec{k}_{g\perp},
\label{16}
\end{equation}
The  energy  denominators  corresponding  to   dashed   lines   on
diagrams  behave at $P\rightarrow \infty$ and $q_{\perp}\rightarrow
\infty$ as

\begin{equation}
\frac{1}{2E_{1}(E+q_{0}- \sum_{i}E_{i})}=\mp
\frac{x}{q_{\perp}^{2}(x_{1}-y_{1})}.
\label{17}
\end{equation}
The upper sign in (16) correspons to fig.3a, and  bottom  sign
to fig.3b diagrams, respectively.
The quark-gluon vertices  on  these diagrams  behave  as

\begin{equation}
\bar{u}(p'_{1})\gamma_{i} u(k'_{1})=q_{i}(x_{1}+y_{1})
+i\sigma_{3} \epsilon_{ik}q_{k}(x_{1}-y_{1})
\label{18}
\end{equation}
for transverse gluons and these  diagrams do  contribute  to
$g_{2}(x)$ (but  do  not  contribute  to  other  structure  functions).
The  diagrams with $q\bar{q}$ - pairs creations or annihilations
(figs. 2c, 2d) do not contain necessary structure
$\epsilon_{ij}q_{j}\vec{s} \vec{q}$ and do not  contribute  to
$g_{2}(x)$. It is easy to understand that nucleon and gluon  spins
 must be aligned in the same direction and the  nucleon  spin-flip  takes
place as it is demonstrated on the diagrams.
   We find the following result for these diagrams:

$$2Mg_{2}^{(2)}(x)=\sqrt{2x_{r}y_{r}} \sqrt{4\pi \alpha_{s}} \sum_{r}
\int d \Gamma^{n} [\delta(x-x_{r})-\delta(x-y_{r})]
\frac{dy_{g}d\vec{k}_{g\perp}}{(2\pi)^{3}2y_{r}y_{g}^{2}}\times$$

\begin{equation}
\sum_{s_{1},...s_{n}}
(\Psi_{s_{i}}^{\ast \downarrow}(x_{i},\vec{p}_{i\perp})
\Psi_{s_{i}\uparrow}^{\uparrow}(x_{i},\vec{p}_{i\perp})+
\Psi_{s_{i}\downarrow}^{\ast \downarrow}(x_{i},\vec{p}_{i\perp})
\Psi_{s_{i}}^{\uparrow}(x_{i},\vec{p}_{i\perp}))
2s_{r}Q_{r}^{2},
\label{19}
\end{equation}
Here $\Psi_{s_{i}\lambda}$=$ 1/2\lambda_{a} \psi_{s_{i}\lambda}^{a}$
($\lambda_{a}$ denotes Gell-Mann matrices) is the wave function of
nucleon consisting of n-partons   and  one  transverse
gluon with helicity $\lambda$. The Burkhardt-Cottingham  sum  rule  (2) is
fulfilled for $g_{2}(x)$ due to cancelation of  the  contributions  of
fig.3a and fig.3b diagrams according to (17).

   Thus, the structure  function  $g_{2}(x)$ in the parton model is
determined by two physically independent mechanisms  which  do  not
contribute to other structure functions, so measuring $g_{2}(x)$ we can
get essentially new information  about quark-gluon structure of the
nucleon,  namely  information  about   angular   orbital   momentum
distribution  and  information  about  gluon  distribution  in  the
nucleon.

   The direct  connection  of  our  results  with  the  results  of
operator product expansion \cite{5,7} (see, also, Ref.10 and  references
therein)
is not obvious and  needs  further  considerations. Note only, that as
it easy to understand from Egs.11,15 $g_{2}^{(1)}(x)$ corresponds to
twist-2 and twist-3  contributions,  whereas $g_{2}^{(2)}(x)$ corresponds
only to twist-3 contribution. The QCD evolution
of $g_{2}^{(2)}(x)$ is,  apparently,  more  complicated  and  differs  from
evolution of $g_{2}^{(1)}(x)$, so the  structure function $g_{2}(x)$
cannot  be evolved from its value at some low $Q^{2}$ , in accordance
with  results of operators product expansion \cite{6,7}.

 \section{Numerical calculations of $ g_{1}(x)$ and $ g_{2}(x)$ in
relativistic quark  model}

In order to estimate the role of twist-3 contribution and to
understand in details the physical interpretation of $ g_{2}(x)$ it is
worth to calculate this function in a specific model. In this section the
results of such calculation  in the relativistic quark model (RQM) of Ref.6 are
presented. We actually assume that at low virtualities nucleon consists
of three valence constituent quarks only and do not discuss the
contributions of diagrams of fig.3. In such a simple model
it is impossible to reproduce the the correct Regge behavior at
$x\rightarrow 1$.  So, our results are expected to be reasonable in the
range of not small values of $x$ ($x\leq 0.2-0.3$), where the contributions
of nonperturbative quark-antiquark
pairs are expected to be small. The results obtained must be evaluated
 into the range of experimental values of $Q^{2}$ . In general, it is
impossible to write down the simple evolution equation for twist-3 contribution
to
$g_{2}$(x) which would allow to connect this function at different values of
$Q^{2}$
\cite{10,11}.
Nevertheless in Ref.12 it was found that such approximate equation
(which becames exact in $N\rightarrow \infty$ limit) exist and
utilizing this result it is possible to evaluate our results into
the range of experimental momentum transfer.

\begin{figure}[htb]  \centering \unitlength 1mm
\begin{picture}(50,97)
\put(-35,100){\begin{picture}(50,97)
\includegraphics{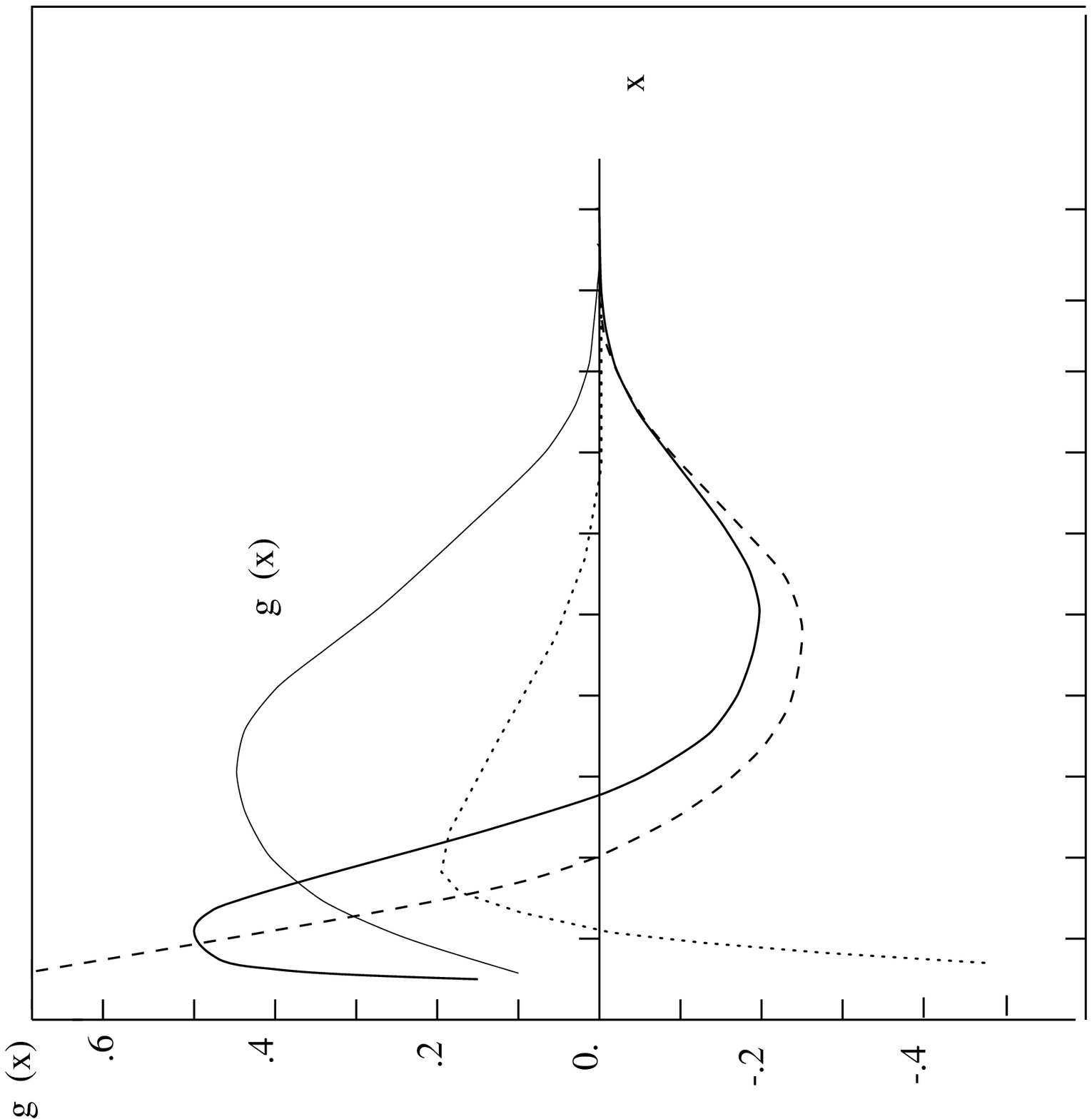}
\end{picture}}
\end{picture}
\caption{\captionsize The results of numerical calculations of $g_{2}(x)$ and
      $g_{2}(x)$ in relativistic quark model.}
\label{pis}
\end{figure}

 In RQM the spin dependent structure functions $ g_{1}(x)$  and
$ g_{2}(x)$ are
determined by Egs.11,15 with n=3. The nucleon wave functions (13)
in the IMF have following form \cite{6} \footnote{Different variants of
relativization of quark model differ, in general, by explicit form of
Melosh matrices, see e.g.,\cite{13} and references therein.}:

\begin {equation}
\Psi_{s_{1}s_{2}s_{3}}^{s}(x_{i},p_{i\perp})=
\Phi(M_{0}^{2})U_{s_{1}s'_{1}}(x_{1},p_{1\perp})
U_{s_{2}s'_{2}}(x_{2},p_{2\perp})U_{s_{3}s'_{3}}(x_{3},p_{3\perp})
\chi_{s'_{1}s'_{2}s'_{3}}^{s}
\label{20}
\end {equation}
where $\Phi(M_{0}^{2})$ is a radial part which supposed to depend only of
one argument - invariant mass of the system of quarks, composing nucleon
\cite{6},

\begin {equation}
M_{0}^{2}=\sum_{i=1}^{3}\frac{p_{i\perp}^{2}+m^{2}}{x_{i}},
\label{b}
\end {equation}
where $m$ stands for constituent quark mass; $\chi_{s'_{1}s'_{2}s'_{3}}^{s}$
is the spin orbital part of wave function which is supposed to coincide
with nonrelativistic wave functions of naive quark model. The Melosh matrices,

\begin {equation}
U(x_{i},p_{i\perp})=\frac{m+M_{0}x_{i}+i\epsilon_{mn}\sigma_{m}
(p_{i\perp})_{n}}{(m+M_{0}x_{i})^{2}+p_{i\perp}^{2}}
\label{mel}
\end {equation}
determine the transformation of $\chi_{s_{1}s_{2}s_{3}}^{s}$
functions from rest frame into IMF.  In Ref.6 a good description of nucleon
static parameters were obtained under assumption that nucleon wave function
is pure $[56.0^{+}]$ representation of SU(6) group. For radial wave
function it was supposed the following form:

\begin {equation}
\Phi(M_{0}^{2})=N\exp(-M_{0}^{2}/6\alpha^{2})
\label{fi}
\end {equation}
The fitting parameters, $\alpha$ and $m$ turned out to be \cite{6}:

\begin {equation}
\alpha=380\pm60Mev, m=270\pm30Mev
\label{alph}
\end {equation}
Note, that parameter $\alpha$ characterizes the mean square momentum of
quark in nucleon: $p_{\perp}^{2}\simeq2/3\alpha^{2}$. It is very important
to note that due to Melosh transformation in \ref{20} the nucleon IMF wave
function depends on quark transvers momentum even for pure $[56.0^{-}]$
representation. This actually means that that due to Melosh transformation in
pure SU(6) wave function in IMF arises the admixture of states with nonzero
value of angular orbital momentum. It makes it possible to derive the nonzero
value of $g_{2}(x)$. We find the following results for spin dependent
structure functions:

\begin {equation}
2g_{1}^{p}(x)=\frac{4Q_{u}^{2}-Q_{d}^{2}}{3}\int{d\Gamma^{3}\delta(x-x_{1})[1-\frac
{2p_{1\perp}^{2}}{(m+M_{0}x_{1})^{2}+p_{1\perp}^{2}}]}
\label{g1}
\end{equation}

\begin {equation}
2Mg_{2}^{p}(x)=\frac{4Q_{u}^{2}-Q_{d}^{2}}{3}\frac{d}{dx}\int{d\Gamma^{3}\delta(x-x_{1})
\frac{p_{1\perp}^{2}(m+M_{0}x_{1})}{(m+M_{0}x_{1})^{2}+p_{1\perp}^{2}}}
\label{g2}
\end{equation}
For the neutron structure functions the replacement
$Q_{u}\leftrightarrow Q_{d}$ must be inserted in (\ref{g1},\ref{g2}) and we
find:

\begin {equation}
g_{1}^{n}(x)=g_{2}^{n}(x)=0
\label{gg12}
\end{equation}
This result qualitively agrees with low experimental value of
 $g_{1}^{n}(x)$ . The nonzero values of $g_{1}^{n}(x)$ and $g_{2}^{n}(x)$
can be
obtained if we take into account the admixture of high multiplets in the
nucleon wave function.

The nucleon $\beta$ -decay coupling

\begin{equation}
g_{A}=6\int{(g_{1}^{p}(x)-g_{1}^{n}(x))dx}\simeq 1.22
\label{ga}
\end{equation}
is well described in the model.

The part of nucleon spin carried by quarks in IMF is given by

\begin{equation}
2\langle{S_{z}}\rangle =\int{d\Gamma^{3}[1-\frac
{2p_{1\perp}^{2}}{(m+M_{0}x_{1})^{2}+p_{1\perp}^{2}}]}\simeq0.73
\label{sz}
\end{equation}
and turned out to be less than unity even for pure SU(6) nucleon wave
function (see also Ref.14). The missing part of nucleon spin is carried
by angular orbital momentum $L_{z}$  which arise due to Melosh
transformations. It is easy to check that $\langle{L_{z}}\rangle$ exactly
correspond to
the second term in square brackets in (30). But this interesting relativistic
effect cannot account for the spin deficit which follows from
experimental data. Keeping in mind these results (29-30) it is natural to
assume that the considered model more succesefully can be applied in flavour
non-singlet channel. This is, apparently, connected with possible
cancelation of unknown contributions which could be essential in
singlet channel. So, for comparison with expirement or with predictions
of other models it is more reasonable to consider the differences of
proton and neutron structure functions.

The results of numerical calculations are presented on Fig.4. The
twist-3 contribution to $g_{2}(x)$ was obtained by substructing from
$g_{2}(x)$ (27) the twist-2 contribution which is determined by $g_{1}(x)$
\cite{15}:

\begin{equation}
g_{2}^{tw.2}(x)=-g_{1}(x)+\int_{x}^{1}{\frac{g_{1}(y)}{y}dy}
\label{ww}
\end{equation}

Our results are very close to the bag model calculations of Ref.7 and
indicate that twist-3 contribution is not small. Remind that these
results correspond to low resolution scale, of order of typical
constituent quark virtualities in the nucleon. The evolution to the
experimental values of $Q^{2}$ will increase our predictions considerably
(about 5-8 times) in the range of $x\geq 0.5$.

The second moments of twist-3 contribution to $g_{2}(x$) were calculated
in Refs.8,9 in framework of QCD sum rules:

\begin{equation}
M_{p-n}^{(2)}=\int{(g_{2}^{p}-g_{2}^{n})x^{2}dx}=0.008\pm0.004
\label{bbk}
\end{equation}

\begin{equation}
M_{p+n}^{(2)}=\int{(g_{2}^{p}+g_{2}^{n})x^{2}dx}=-0.009\pm0.004
\label{bbk1}
\end{equation}

Our result for difference $g_{2}^{p}(x) -g_{2}^{n}(x)$
\begin{equation}
M_{p-n}^{(2)}\simeq 0.002
\label{bbk2}
\end{equation}
agrees with (31) by sign and by order of magnitude.

\section{Conclusion}

In the special infinite momentum frame the structure function  $g_{2}(x)$
has a reasonable interpretation in terms of nucleon parton wave function
(but not in terms of probabilities). We have two contribution which are
phenomenologically independent. The  first contribution corresponds to Compton
diagrams with different values of angular orbital momenta projections in the
initial and final nucleon wave functions. This  contribution is proportional
to quark transvers momentum and corresponds to
twist-2 and twist-3 contributions in terms of OPE. The second
contribution is determined by Compton diagrams with different number of
gluons in initial and final nucleon wave functions and correspond only to
twist-3.

     According to independent estimates in different models the function
$g_{2}(x)$ and the twist-3 contribution to  $g_{2}(x)$, which really contains
new information on quark-gluon interaction in nucleon are not small. So,
the projected measurements of this function are reasonable and will provide
new information on nucleon structure.
\section{Acnowledgements}
I would like to thank E.M.Levin for stimulating discussion at the
beginning of the work and V.M.Braun, who brought my attention to the
consideration of $g_{2}(x)$ structure function in the relativistic quark model
and for the discussion of the results. The last part of this work was
completed at DESY and I would like to thank  DESY Hermes group for
kind hospitality.

This work was supported by International Science Foundation,
Grant $\sharp$ RYE OOO and by the Project INTAS 93-283.

\end{document}